\documentclass[9pt]{elife}
\usepackage{lipsum} 
\usepackage[version=4]{mhchem}
\usepackage{siunitx}
\DeclareSIUnit\Molar{M}

\title{Non-genetic inheritance restraint of cell-to-cell variation}

\author[1]{Harsh Vashistha}
\author[1]{Maryam Kohram}
\author[1,2*]{Hanna Salman}
\affil[1]{Department of Physics and Astronomy, The Dietrich School of Arts and Sciences, University of Pittsburgh, Pittsburgh, PA}
\affil[2]{Department of Computational and Systems Biology, School of Medicine, University of Pittsburgh, Pittsburgh, PA
}

\corr{hsalman@pitt.edu}{HS}
\begin{document}

\maketitle
\begin{abstract}
Heterogeneity in physical and functional characteristics of cells (e.g. size, cycle time, growth rate, protein concentration) proliferates within an isogenic population due to stochasticity in intracellular biochemical processes and in the distribution of resources during divisions. Conversely, it is limited in part by the inheritance of cellular components between consecutive generations. Here we introduce a new experimental method for measuring proliferation of heterogeneity in bacterial cell characteristics, based on measuring how two sister cells become different from each other over time. Our measurements provide the inheritance dynamics of different cellular properties, and the “inertia” of cells to maintain these properties along time. We find that inheritance dynamics are property- specific, and can exhibit long-term memory ($\sim$10 generations) that works to restrain variation among cells. Our results can reveal mechanisms of non-genetic inheritance in bacteria and help understand how cells control their properties and heterogeneity within isogenic cell populations.
\end{abstract}


\section{Introduction}

One of the main challenges in biological physics today is to quantitatively predict the change over time in cells’ physical and functional characteristics, such as cell size, growth rate, cell-cycle time, and gene expression. All cellular characteristics are determined at all times by the interaction of genetic and non-genetic factors. While genetic information passed from generation to the next is the main scheme, by which cells conserve their characteristics, non-genetic cellular components, such as all proteins, RNA and other chemicals, are also transferred between consecutive generations and thus influence the state of the cell's characteristics (or its phenotype) in future generations (\cite{Lambert2014}; \cite{Robert2010}). The mechanism of genetic information transfer between generations, as well as how this information is expressed, are mostly understood (\cite{Casadesus2006}; \cite{Chen2017}; \cite{Turnbough2019}). This information can be altered by rare occurring processes such as mutations, lateral gene transfer, or gene loss (\cite{Bryant2012}; \cite{Robert2018}). Therefore, changes resulting from genetic alterations emerge over very long timescales (several 10s of generations). On the other hand, inheritance of non-genetic cellular components, which are subject to a considerable level of fluctuations, can influence cellular characteristics at shorter timescales (\cite{Casadesus2013}; \cite{Huh2011}; \cite{Norman2013}; \cite{Veening2008}).

Here we focus on understanding how robust cellular characteristics are to intrinsic sources (stochastic gene expression and division noise) and extrinsic sources (environmental fluctuations) of variation, and how cells that emerge from a single mother develop distinct features and over what time scale. While our understanding of variation sources has increased significantly over the past two decades (\cite{Ackermann2015}; \cite{Avery2006}; \cite{Elowitz2002}), progress in understanding non-genetic inheritance and its contribution to restraining the proliferation of heterogeneity has been extremely limited. Extensive studies have been dedicated to revealing the different non-genetic mechanisms that influence specific cellular processes and how they are inherited over time (\cite{Chai2010}; \cite{Govers2017}; \cite{Mosheiff2018}; \cite{Sandler2015}; \cite{Wakamoto2005}). However, the cell's phenotype is determined by the integration of multiple processes. Thus, to predict the inheritance dynamics of a cellular phenotype, we need to measure the inheritance dynamics directly rather than characterizing the effect of individual inheritance mechanisms separately. Progress in this research has been drastically hindered by the limited experimental techniques that can provide reliable quantitative measurements.

The recent development of the “mother machine” (\cite{Brenner2015}; \cite{Wang2010}), has provided valuable data of growth and division, as well as protein expression dynamics. These data have been used to gain insight into non-genetic inheritance and cellular memory. The results obtained have consistently showed that non-genetic memory in bacteria is almost completely erased within two generation (\cite{Susman2018}; \cite{Tanouchi2015}; \cite{Wang2010}). This has been also the conclusion of theoretical calculations of cell size autocorrelation (\cite{Ho2018}; \cite{Susman2018}), which are based on the adder model (\cite{Amir2014}; \cite{Taheri-Araghi2015}; \cite{SI20191760}) for size homeostasis. The consensus of previous experimental studies is founded on the calculation of the autocorrelation function (ACF) for the different measurable cellular properties, such as cell size, growth rate, cell cycle time, and protein content. It is important to note that in calculating the ACF, measurements of cells from different traps of the mother machine are averaged together. However, small variation in the traps sizes can manifest during the fabrication process, which can lead to distinct environments in different traps (\cite{Yang2018}). In addition, cells might experience slightly different environments at different times resulting from thermal fluctuations and their dynamic interaction with their surroundings, i.e. environmental fluctuations can influence the cell's growth and division dynamics, which in turn can change the cell's micro-environment through consumption of nutrients and/or secretion of other chemicals. As a result of the individuality of the cell-environment interaction, different microniches can be created in different traps (Figure1-Figure supplement 1A), giving rise to diverse patterns of growth and division dynamics and therefore distinct ACFs (Figure1-Figure supplement 1B) (\cite{Susman2018}; \cite{Yang2018}; \cite{Tanouchi2015}). Averaging over many traps, with such various ACFs, will thus erases the dynamics of cellular memory.

 To overcome this hurdle, we have developed a new measurement technique, which enables us to separate environmental effects from cellular ones. The technique is based on a new microfluidic device that allows trapping two cells immediately after they divide from a single mother simultaneously, and sustain them right next to each other for extended time. Thus, with this technique we track the lineages of the two sister cells from the time of their birth and follow them as they age together for tens of generations. This enables us to measure how two cells that originate from the same mother become different over time, while experiencing exactly the same environment. Thus, we are able to measure the non-genetic memory of bacterial cells for several different traits. Our results reveal important features of cellular memory. We find that different traits of the cell exhibit different memory patterns with distinct timescales. While the cell cycle time and cell size exhibit slow exponential decay of their memory that extends over several generations, other cellular features exhibit complex memory dynamics over time. The growth rates of two sister cells, for example, diverge immediately after division, but re-converge towards the end of the first cell cycle and subsequently persist together for several generations. In comparison, the mean fluorescence intensities, reporting gene expression, are identical in both cells immediately after they separate but diverge within two cell cycles.

\section{Results}

Our new microfluidic device, dubbed the "sisters machine" (Figure 1A), consists of 30$\mu$m long narrow trapping channels (1$\mu$m × 1$\mu$m) open at one end to a wide channel (30$\mu$m × 30$\mu$m), through which fresh medium is continuously pumped to supply nutrients to cells in the traps and wash away cells that are pushed out of them. Here however, every two neighboring trapping channels are joined on the closed end through a v-shaped connection of the same width and height. The tip of the v-shaped connection is made 0.5$\mu$m narrower than the rest of the channel to reduce the likelihood of cells passing from one side to the other (Figure 1B). Therefore, once it happens, the cells at the tip will remain there, while we track their growth and division events, and measure their size and protein expression (Figure 1C and D), until the next cell passage occurs, which can take 10s of generations (see Supplementary movie). The environment in this setup is identical for both cells at the tip of the v-shaped connection, as they are kept in close proximity to each other. This ensures that differences observed between the two cells are due to internal cellular factors only. Note as well, that division in this channel does not alter the statistics of sister cells (SCs') relative sizes, growth rates, or generation times, in comparison to that observed in the case of division in straight channels (Figure1-Figure supplement 2).

\begin{figure}

\includegraphics[width=14cm]{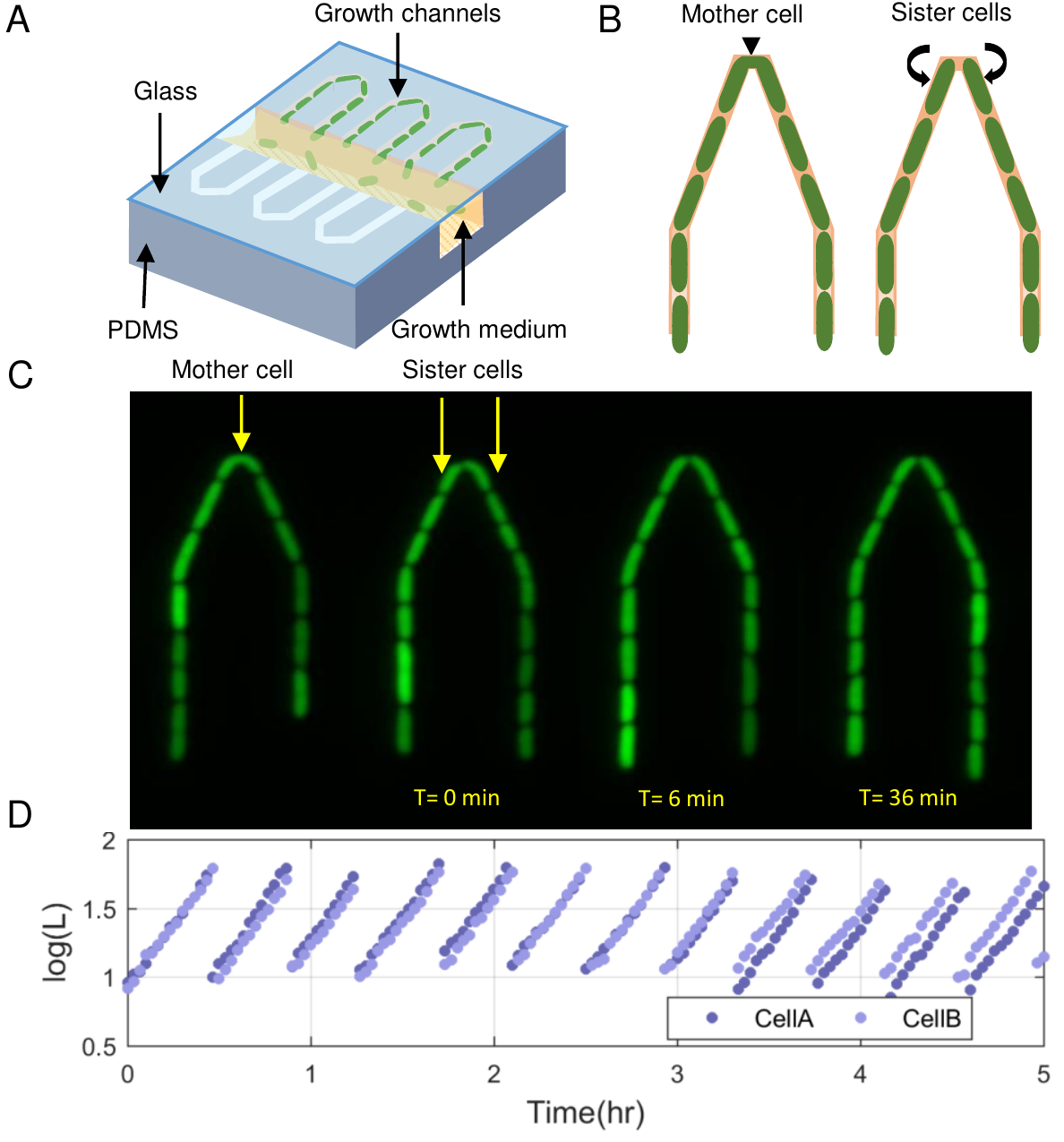}
\caption{Scheme of the experimental setup for tracking sister cells. (A) Long (30$\mu$m) narrow traps (1$\mu$m × 1$\mu$m) are connected on one end and open on the other to wide (30$\mu$m × 30$\mu$m) perpendicular flow channels through which fresh medium is pumped and washes out cells that are pushed out of the traps. (B) Illustration of SCs being born from a single mother cell at the tip of the trap, as can be also seen in real fluorescence images of the cells in the trap (C), which are then followed for a long time (see Supplementary movie). (D) Section of example traces of two sister cells from the time they are born, which shows how they become different over time.}

\figsupp[Individuality of cellular growth dynamics in different microenvironments.]{(A) Probability Distribution Function (PDF) of the absolute difference in the average growth rate of two sister cells (SCs) is compared with absolute difference in the average growth rate of two randomly paired cells (RPs) growing in separate traps in the same device.  The standard deviation of the  difference for SCs ($ \sigma_{SCs}$) is almost half of the  calculated value for RPs ($\sigma_{RPs} $). This shows that cells grow with different average growth rate in different traps and supports the idea of formation of microniches in the microfluidic device. (Inset) Depicts the cell length of two pairs of SCs measured in two different V-Shaped traps as a function of time. The length of each cell is presented in a "stitched" form, where the length of the cell in each cell cycle is adjusted to start from the length of the cell at the end of the previous cycle, ignoring by this the division events. This is done by dividing the length in each cycle by the starting length and multiplying it by the length of the cell at the end of the previous cycle. This presentation emphasizes the difference in the average growth rates measured in different traps. Note however, that each pair of SCs exhibits similar average growth rate. (B) The ACFs of individual lineages, measured in the same experiment in separate traps in the mother machine, are presented in different colors. Each ACF was calculated from a lineage longer than 150 generations to maximize the statistics. Note that each ACF exhibits distinct dynamical pattern. Averaging all ACFs results in a simple exponentially decaying function with a decay time of $\sim$2 generations depicted by the black line in the graph.  }{\includegraphics[width=1.0 \linewidth]{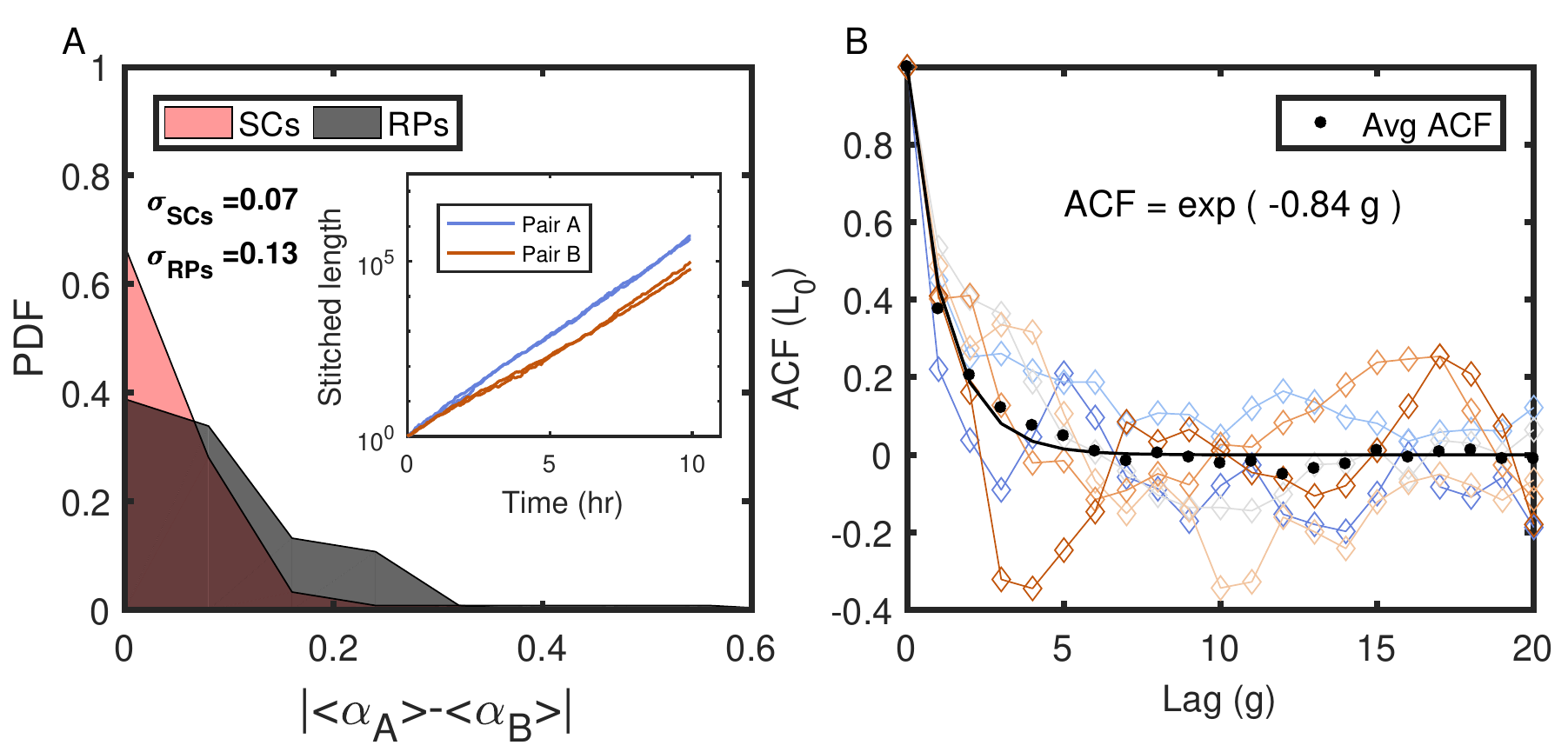}
}

\figsupp[The effect of the v-shaped channel on the distribution of the different cellular characteristics between SCs during division.]{(A) Probability Distribution function (PDF) of the difference in the first cell cycle time of two sister cells after separation relative to the population's average cycle time under the same experimental conditions. (B) PDF of the difference in cell length between the sister cells immediately after division relative to the population's average length at the start of the cell cycle. (C) PDF of the difference in the growth rate of the two sister cells after separation relative to the population's average growth rate. The difference measured in the straight channels here is larger than that measured in the v-shaped channels. This could be due to the fact that the two cells in the mother machine trap are at different distance from the nutrients diffusing from the flow channel into the traps. This has been shown before to result in variation in the cells growth rate \cite{Yang2018}. In all graphs the blue curves represent the distributions measured in our new device with the v-shaped channels using 194 pairs, while the brown curves were measured in the straight channels of the mother machine using 198 pairs.  }{\includegraphics[width=1.0 \linewidth]{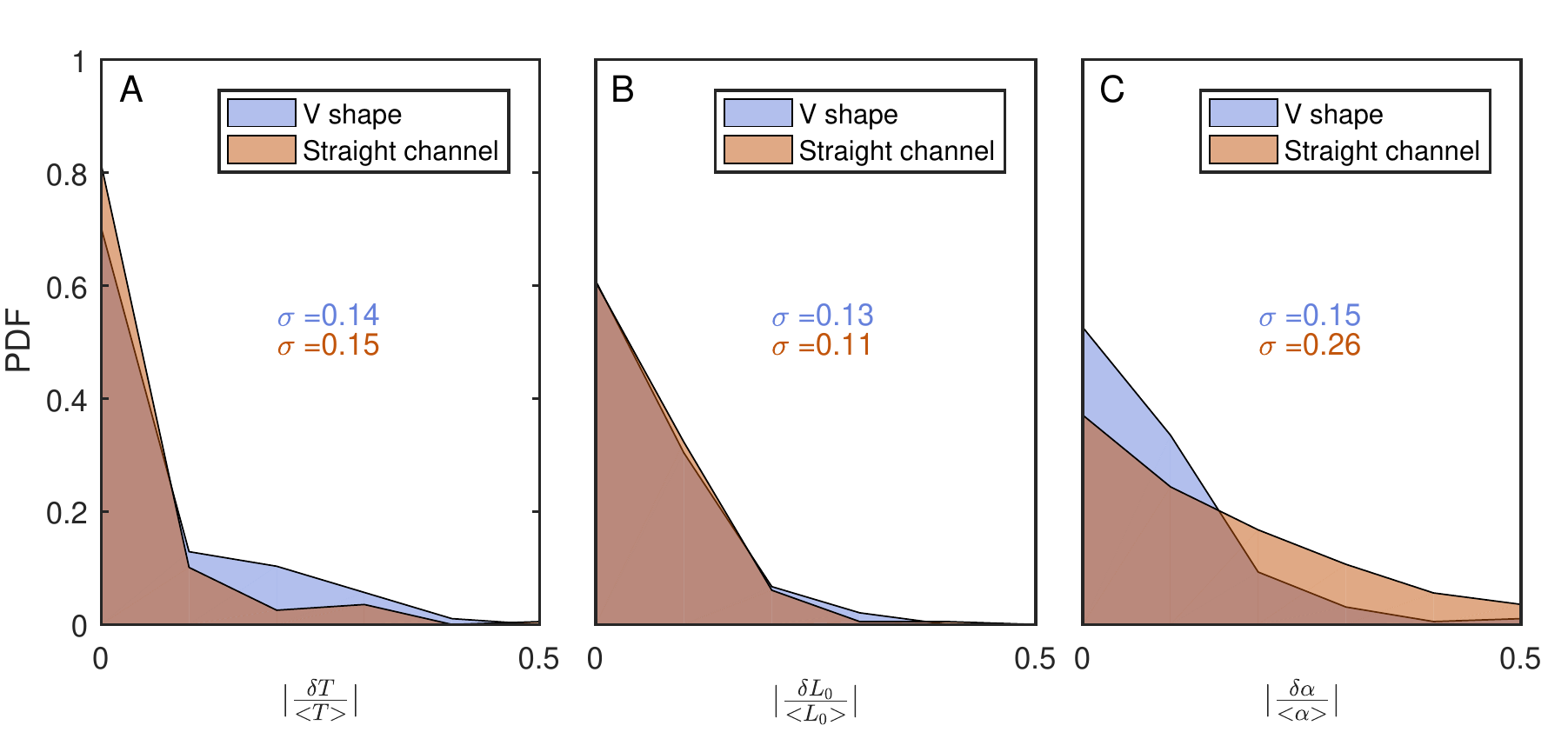}
}

\end{figure}

Using this setup, we successfully trapped pairs of cells next to each other for 20 – 160 generations. Images of the cells in both DIC and fluorescence modes were acquired every 3 minutes. Under our experimental conditions (cells growing in LB medium at 32$^{\circ}$C) the average generation time was ${34\pm7}$ minutes, which provided $ \sim 11 $ images every generation. The acquired images were used to measure various cellular characteristics as a function of time, including cell size, protein concentration, growth rate, and generation time. To measure cellular memory, we replace the ACF, used in previous studies, with the Pearson correlation function (PCF) between pairs of cells:

\begin{equation}
PCF^{(y)}(t)= \frac{1}{\sigma_{y^{(1)}} \sigma_{y^{(2)}}}\sum_{i=1}^{n} (y_i^{(1)} (t)-<y^{(1) } >).(y_i^{(2)} (t)-<y^{(2) } >)
\end{equation}

where y is the cellular property of interest, t is the measurement time, n is the number of cell pairs measured, $\sigma_y$ the population standard deviation of y and (1) and (2) represent the two cells being considered. $PCF^{(y)}(t)$ is therefore a measure of the correlation between the values of a specific cellular property at time t. We use this correlation function to compare three types of cell pairs (Figure 2A): 1) Sister cells (SCs) are cells that originate from the same mother at time 0, and therefore the value of PCF at time 0 is 1. 2) Neighbor cells (NCs) are cells that reside next to each other at the tip of the v-shaped connection. However, NCs are cells that do not originate from the same mother. They are cells that happen to enter into both sides of the same v-shaped channel from the start of the experiment. We initiate their tracking though, only when they happen to divide at the same time, such that at time 0 they are both at the start of a new cell cycle, and if their length is almost identical at that point in time. This choice is to ensure that any long-term correlation measured in SCs does not stem from a size homeostasis mechanism, which would maintain the size of both cells similar for several generations if they start similar. 3) Random cell pairs (RPs) are cells that reside in different traps and their lineages are aligned artificially even though they can be measured at different times. In this case, t is measured relative to the alignment point, which is chosen to be at the start of the cell cycle for both cells. Since NCs and RPs do not originate from the same mother at time 0, the PCF is measured from the first generation only, and we set it to be 1 at time 0. Comparing the correlation of NCs, which experience the same environmental conditions at the same time, with that of RPs allows us to determine the effect of the environment on the correlation. On the other hand, the comparison of SCs with NCs provides the effect of cellular factors (i.e. epigenetics) that are shared between SCs, on the correlation function. This in turn allows us to determine the cellular memory of a specific property resulting from shared information passed on from the mother to the two sisters (see Appendix for the mathematical relationship between the different measures).


\begin{figure}
\includegraphics[width=12cm]{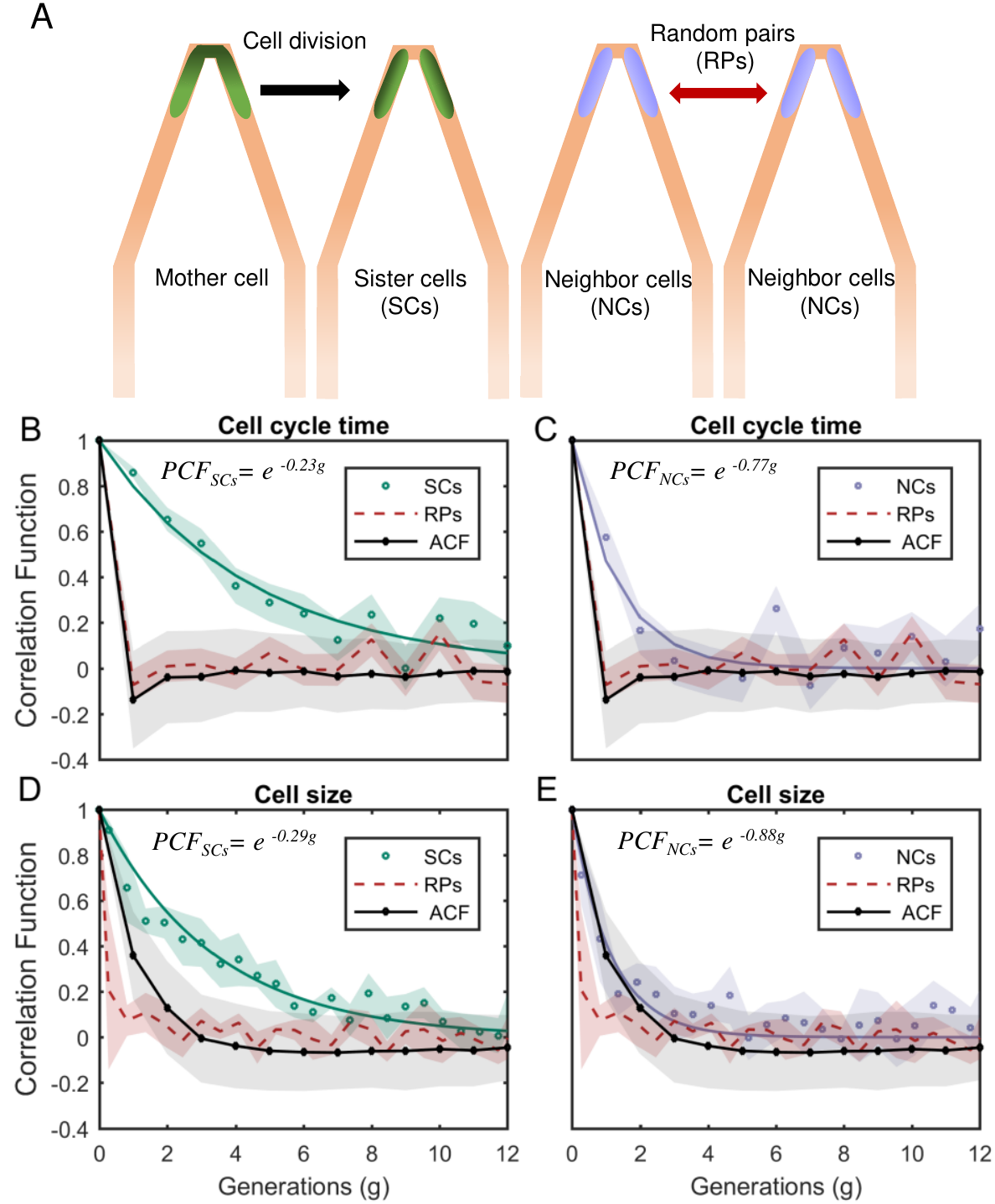}
\caption{PCF of cell cycle time and cell size measured in cell pairs as a function of number of generations. (A) Three types of pairs used for calculating PCF. (B) PCF of cell cycle time for SCs (122 pairs from 3 separate experiments) exhibit memory that extends for almost 9 generations (half lifetime $\sim$ 4.5 generations). This is $\sim$ 3.5x longer than the half lifetime of NCs PCF (calculated using a 100 pairs from 3 separate experiments) (C), which is comparable to the ACF (half lifetime $\sim$ 1 generation). (D) Similarly, SCs exhibit strong cell size correlation that decays slowly over a long time (half lifetime $\sim$3.5 generations), while NCs show almost no correlation in cell size similar to ACF of initial sizes (half lifetime $\sim$1 generation). For details of the cell-cycle time PCF and errors calculation see SI and Figure2–Figure supplement 1 and 2. PCF values for cell size were calculated in similar way to cell-cycle time, and were then averaged over a window of six consecutive time frames (15 minutes time window) (See Figure2–Figure supplement 4 for raw data). Shaded area represents the standard deviation of the average. The equations in the graphs represent the best fit to the PCF depicted in each graph with g is generation number. }

\figsupp[Distributions of different cell parameters.]{ In order to avoid artifacts arising in calculations due to differences between experiments carried out on different days, raw data from these experiments was normalized by subtracting the mean ($\mu$) and dividing by the standard deviation ($\sigma$) for each experiment separately. Later, this normalized data was combined and used for calculating the PCF and variances for different parameters. (A-B) distributions of cell cycle times (T) before and after normalization. (C-D) distributions of elongation rate ($\alpha$) before and after normalization. (E-F) distributions of mean fluorescence intensity (f) before and after normalization..}{\includegraphics[width=1.1 \linewidth]{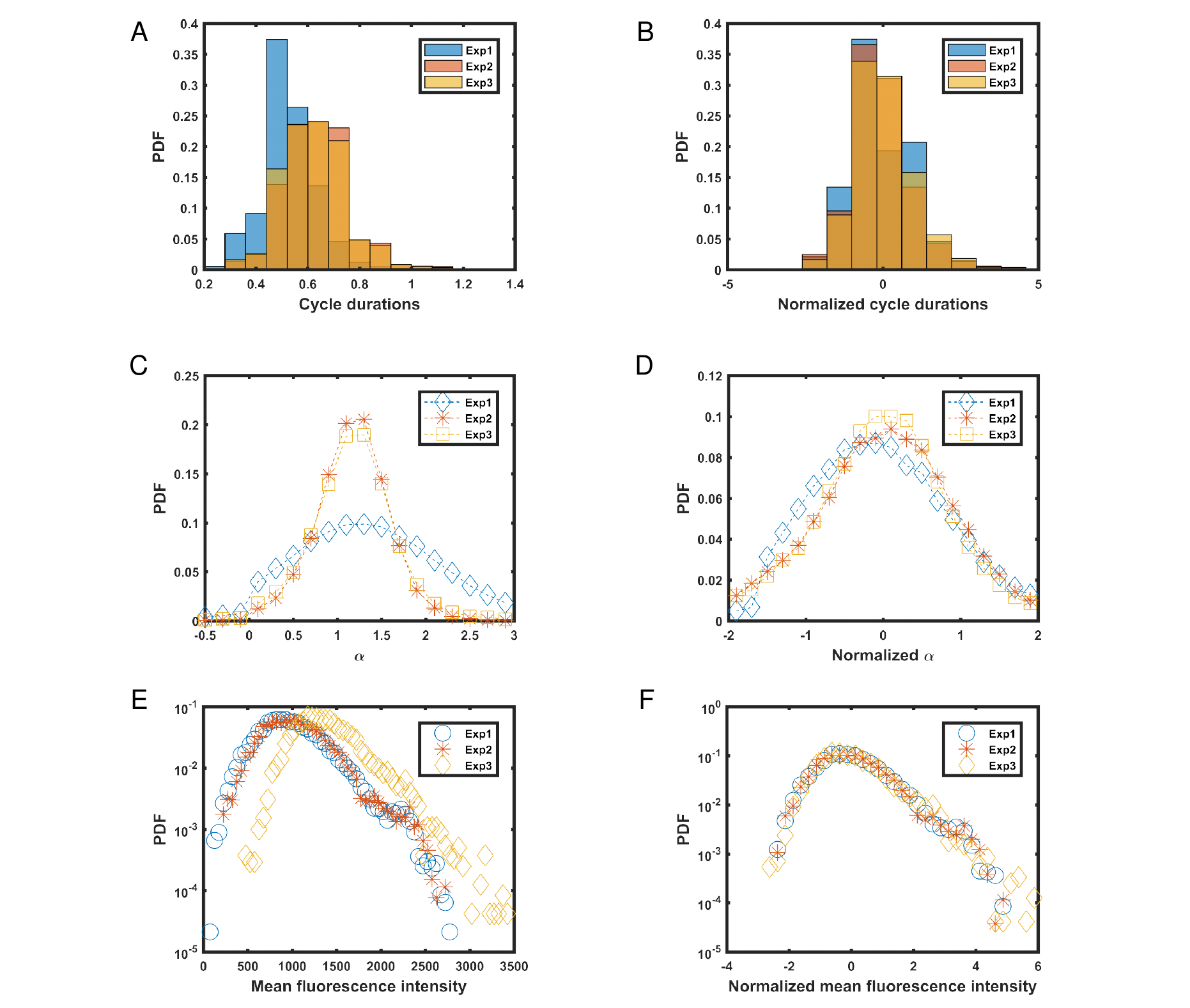}
}

\figsupp[Correlation in cell cycle times for SCs was verified by calculating slopes of best fits to the plots of normalized TimeA vs TimeB.] {(A-I) Slopes of the best fit lines for TimeA vs TimeB show that cell cycle times are strongly correlated for first few generations in SCs. This shows existence of non-genetic memory that restrains the divergence of the phenotypes in cells originating from the same mother cell.}{\includegraphics[width=1.1 \linewidth]{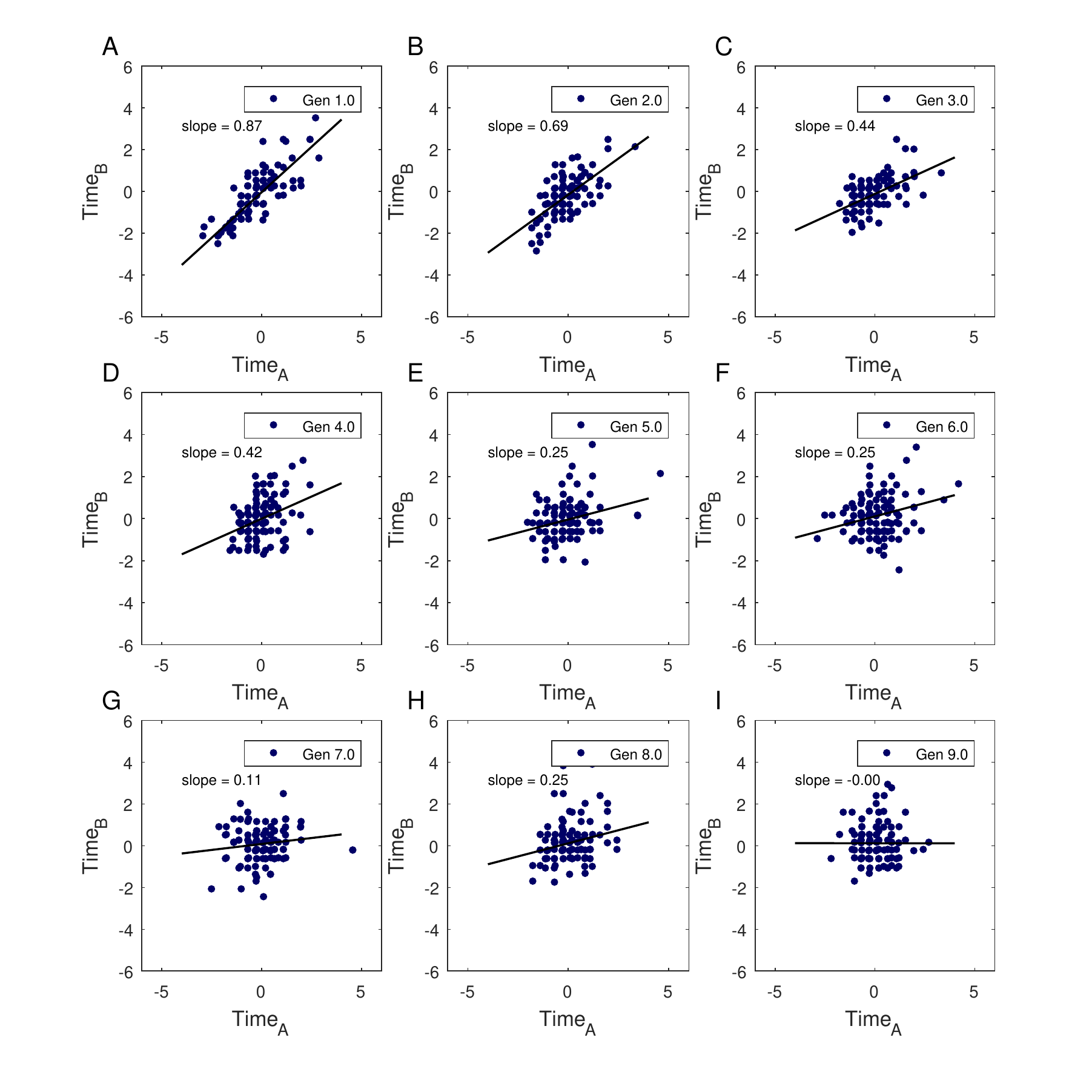}}

\figsupp[The PCF of cell cycle time (T) for SCs in different growth conditions.] {The PCF of SCs cell-cycle time in LB at 37$^{\circ}$ C (57 pairs from 2 separate experiments) (A) and in M9CL at 32$^{\circ}$ C (29 pairs from 2 separate experiments) (B). Existence of strong correlation between cell cycle duration in both (A) and (B) demonstrates the robustness of non-genetic restraint in different experimental conditions. The lines in both graphs are the best fits to the data depicted in the graphs. The decay rate of the correlation in both cases is very similar to that observed in LB medium at 32°C described in the main text (y=exp(-0.23g).  }{\includegraphics[width=14cm]{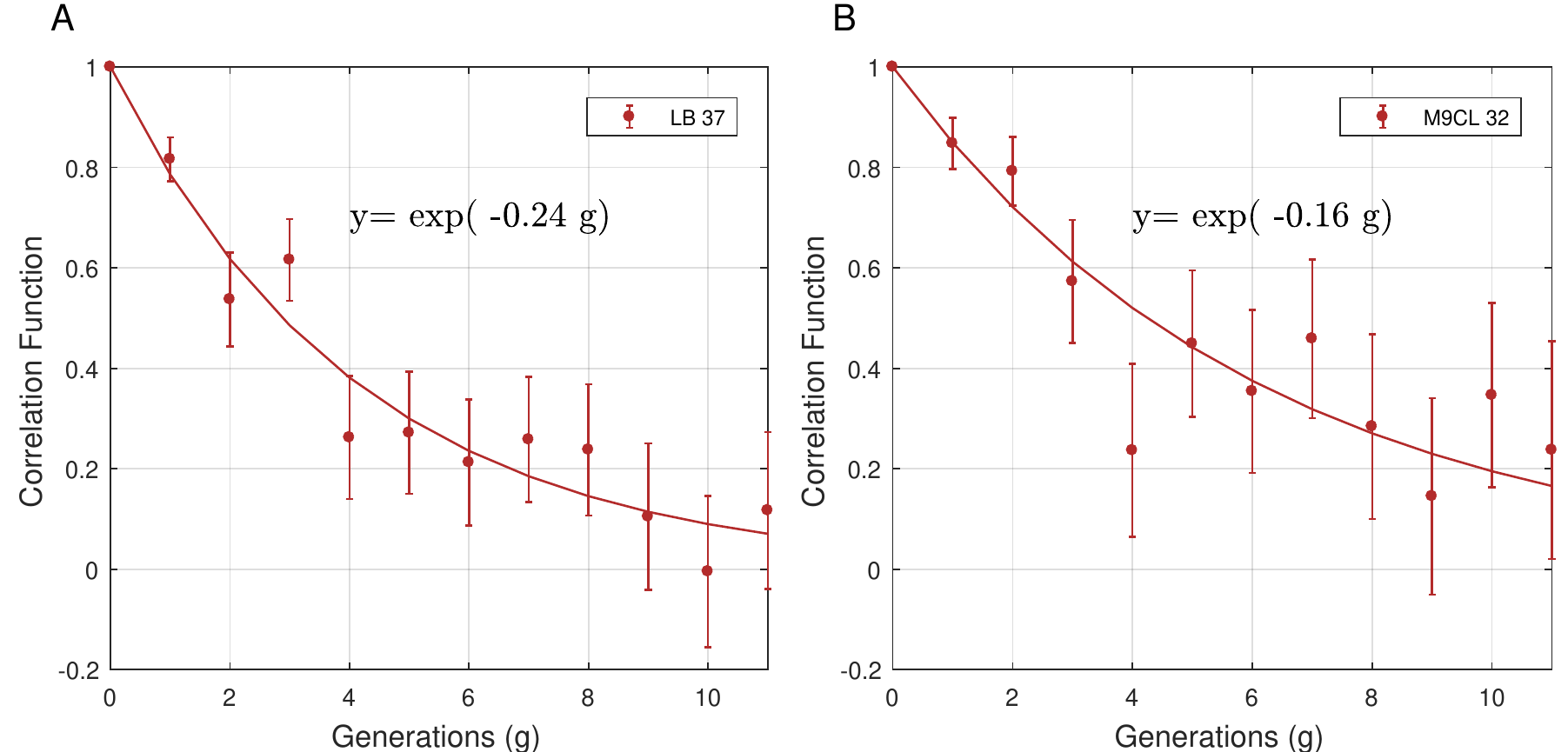}}

\figsupp[Raw PCF values of cell size as a function of time for SCs, NCs and RPs.] {The cell size PCF for SCs (A) and for NCs (B) are compared in both graphs with the cell size ACF and PCF for RPs. Sister cells show strong cell size correlation that decays slowly over a long time. NCs show almost no correlation in cell size similar to ACF of initial sizes. For details of the PCF and errors calculations, refer to earlier SI. }{\includegraphics[width=14cm]{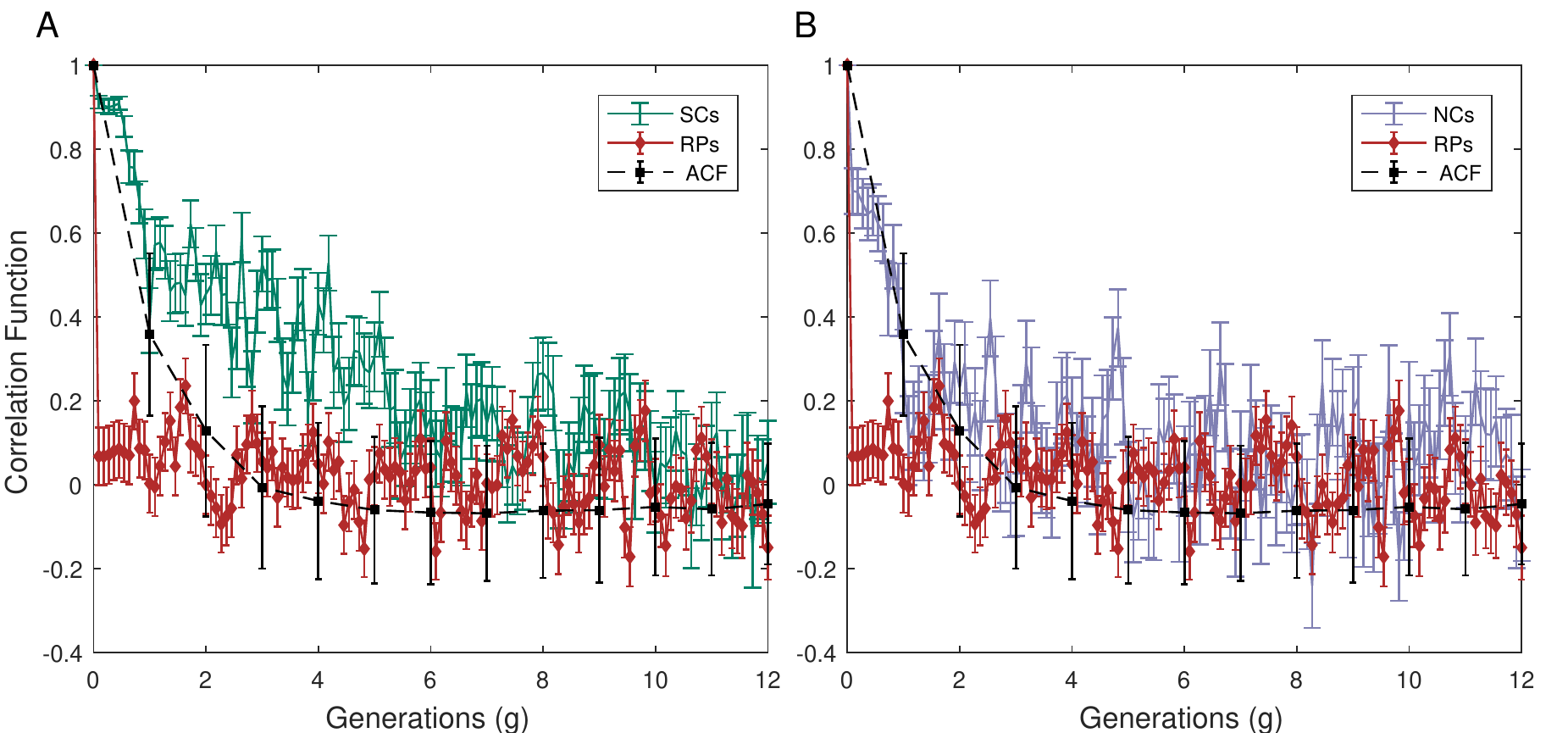}}

\figsupp[PCF values of cell size and cell cycle duration as a function of time for NCs with different starting size.] {PCF of cell-cycle time (A) and cell length (B) for NCs starting from random initial sizes are compared in both graphs with ACF and PCF for RPs. NCs starting with random initial sizes show almost no correlation in cell size or cell cycle time similar to RPs. }{\includegraphics[width=14cm]{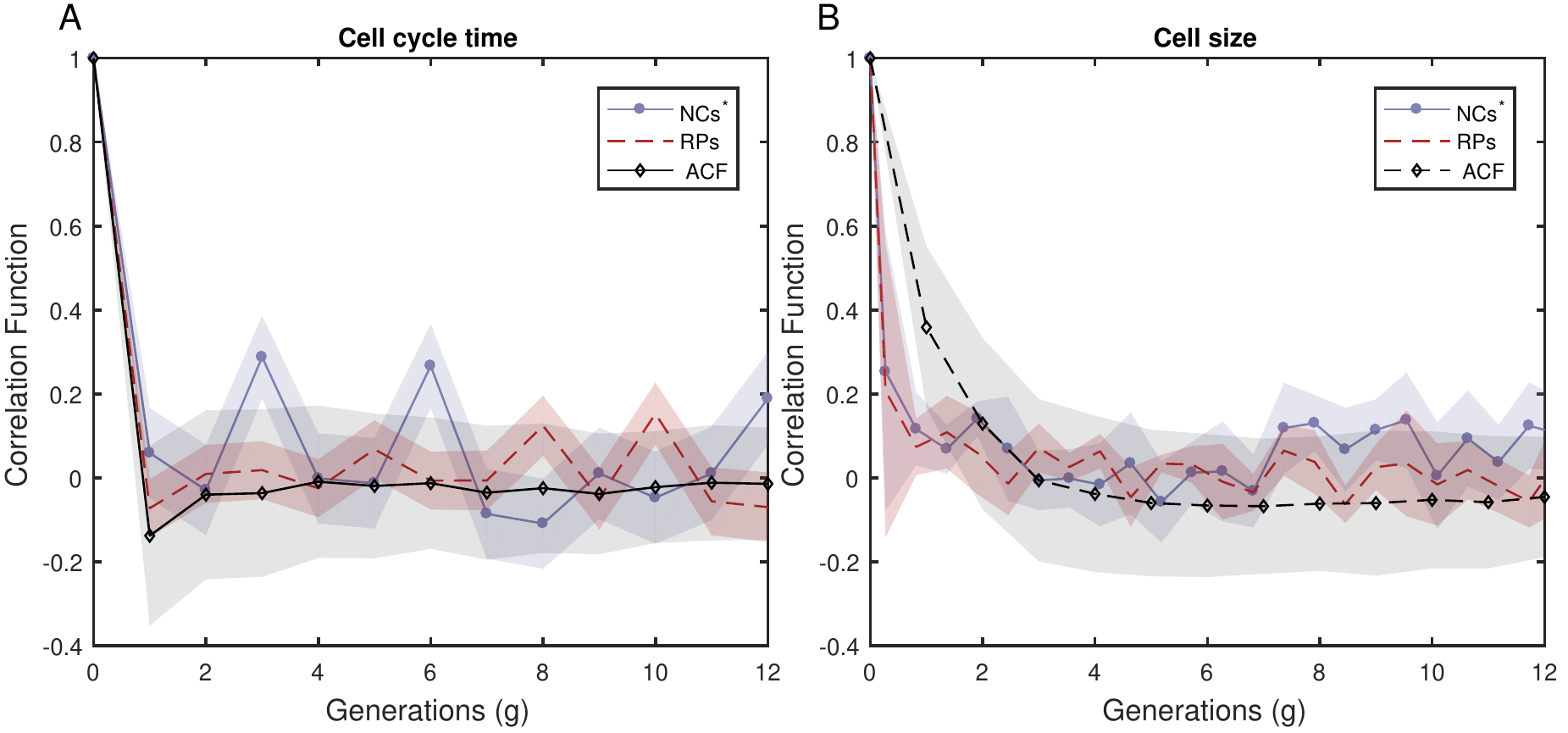}}
\end{figure}

We measured the correlations between the different pair types for cell cycle time (T). We find that T of SCs remain strongly correlated for up to 8 successive cell divisions (Figure 2B also see Figure 2– Figure supplement 1 and 2) regardless of the environmental conditions (Figure 2– Figure supplement 3), while the NCs correlation decays to zero within 3 generation (Figure 2C). These results clearly reveal the effects of epigenetics and environmental conditions on cellular memory when compared to the RPs correlation, which as expected decays to zero within one generation similar to the ACF (Figure 2B and C).

Next, we applied our method to cell size. Also here, our measurements show that SCs correlation decays slowly over $\sim$7 generations (Figure 2D), while the correlation of NCs exhibit fast decay to zero within 2 generations similar to the ACF (Figure 2E). Note that RPs exhibit no correlation from the start of the measurement (Figure 2D and E). These results further demonstrate the existence of strong non-genetic memory that restrains the variability of cell size between SCs for a long time. Unlike the cell cycle time however, the effect of both epigenetic factors and environmental conditions on the cellular memory, appears to extend for a slightly shorter time.

To quantify the increase in variability among cells along time differently, we measured the change in the variance of a cellular property as time advances, which is expected to reach an equilibrium saturation value at long timescales. Measuring how the variance reaches saturation, provides information about cellular memory and the nature of forces acting to restrain variation. The cellular memories of cell cycle time and length, measured using this method, agree well with our previous PCF results (Figure3–Figure supplement 1 and 2). Thus, we have measured the relative fluctuations in the exponential elongation rate of the cell pairs $\delta \alpha$ defined as:
\begin{equation}
\label{eq:CLT}
\delta\alpha(t)=\alpha^{(1) } (t)-\alpha^{(2)} (t)
\end{equation}

where $\alpha (t)=(d \ln{L}/dt)$ is the exponential elongation rate of the cell, L(t) is the cell length at time t, and (1) and (2) distinguish the cell pair (Figure3–Figure supplement 3). As expected, $\delta \alpha$ for all pairs of lineages is randomly distributed with <$\delta \alpha$ >=0 (Figure3–Figure supplement 3), as the elongation rate of all cells fluctuate about a fixed value identical for all cells in the population and depends on the experimental conditions. The variance of $\delta \alpha$ for both RPs ($\sigma_{\delta \alpha_{RPs}}^2$) and NCs ($\sigma_{\delta \alpha_{NCs}}^2$), was found to be constant over time and is similar for both types of cell pairs (Figure 3A). However, the variance of  $\delta \alpha$ for SCs ($\sigma_{\delta \alpha_{SCs}}^2$) exhibits a complex pattern (Figure 3B), which eventually converges to the same value as RPs ($\sigma_{\delta \alpha_{RPs}}^2$) and NCs ($\sigma_{\delta \alpha_{NCs}}^2$). The time it takes for ($\sigma_{\delta \alpha_{SCs}}^2$) to reach saturation extends over almost 8 generations, which again reflects a long memory resulting from epigenetic factors. These results show that, unlike cell cycle time and cell length, elongation rates of SCs immediately after their division from a single mother exhibit the largest variation. This variation decreases to its minimum value within a single cell cycle time ($\sim$30 min). To understand the source of this large variation immediately following separation, we have measured the growth rate over a moving time window of 6 minutes throughout the cell cycle, and compared the results between SCs. Our comparison clearly shows that a SC that receives a smaller size-fraction from its mother exhibits a larger growth rate immediately after division. The growth rate difference between the small and large sisters, decreases to almost zero by the end of the first cell cycle after separation (Figure 3B inset). This result reveals that the exponential growth rate of a cell immediately after division inversely scales with the size-fraction the cell receives from its mother (see also \cite{KOHRAM2020}). It also demonstrates that the difference in the growth rates between SCs changes during the cell cycle indicating that they are not constant throughout the whole cycle as has been accepted so far (\cite{Godin2010}; \cite{Soifer2016}; \cite{Wang2010}).
Note that similar results have been reported recently for \emph{Bacillus subtilis} \cite{NORDHOLT20202238}, and \emph{Caulobacter crescentus} \cite{Banerjee2017}, where it was observed that the growth rate is inversely proportional to the cell size at the start of the cell cycle, and changes as the cell cycle advances.


\begin{figure}
\includegraphics[width=12cm]{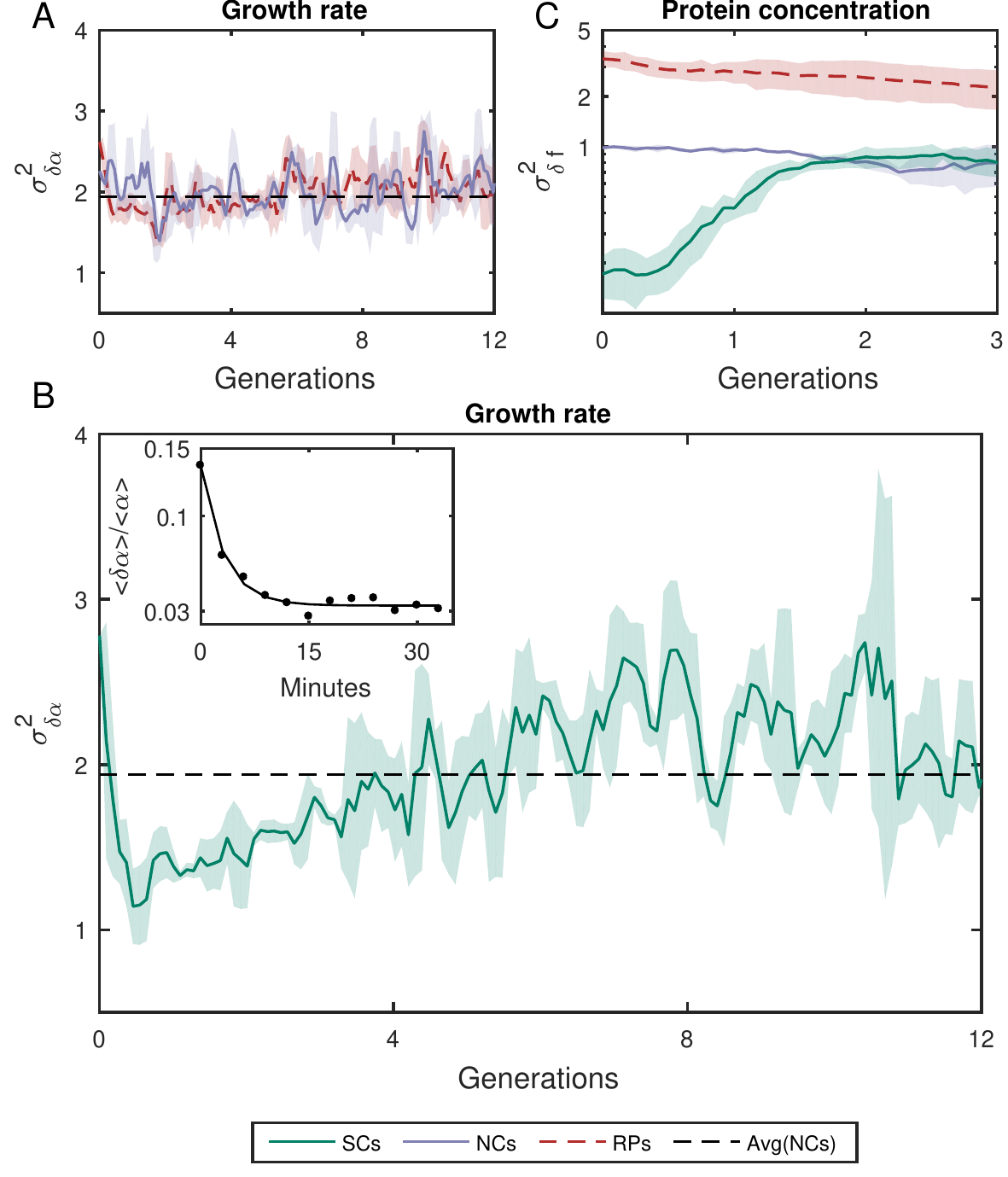}
\caption{Variance  ($\sigma_{\delta \alpha}^2$) as a function of the time. (A) $\sigma^2$ of the growth rate difference ($\delta \alpha$) between cell pairs for NCs and RPs as a function of time (see Figure 3– Figure supplement 3 for the details of the calculation). The variance for both pair types does not change over time. (B) $\delta\alpha$ of SCs, on the other hand, exhibits large variance immediately after separation ($\sim$50percent) higher than NCs and RPs) and rapidly drops to its minimum value within one generation time ($\sim$30 minutes), and increases thereafter for 4 hours ($\sim$8 generations) until saturating at a fixed value equivalent to that observed for NCs and RPs. Each point in a and b is the average over 3 frames moving window, and the shaded area represent the standard deviation of that average. (C) Unlike $\delta \alpha$, $\delta f$ of SCs increases to its saturation value within $\sim$2 generation (see Figure 3– Figure supplement 4 for the details of the calculation). Here, each point represents the average of three different experiments, and the shaded part represents the standard deviation.}

\figsupp[Cell-cycle time variance ($\sigma_{\delta T}^2$) as a function of time.]{ (A-C) Individual traces showing difference in cell cycle times ($\delta T$) for SCs, NCs and RPs respectively. The variance ($\sigma^2$) of cell cycles times differences ($\delta T$)  as a function of time (D) represent the variance of the plots in (A-C) calculated at different time points using $\sigma_{\delta T}^2= <(\delta T)^2>-<\delta T>^2$. $\sigma^2_{\delta T}$ for SCs starts from a small value in first generation and saturate to a constant value after $\sim$7 generations (similar to the timescale obtained from the PCF $\sim$8 generations), while  $\sigma^2_{\delta T}$ for NCs and RPs remain constant over time.}{\includegraphics[width=14cm]{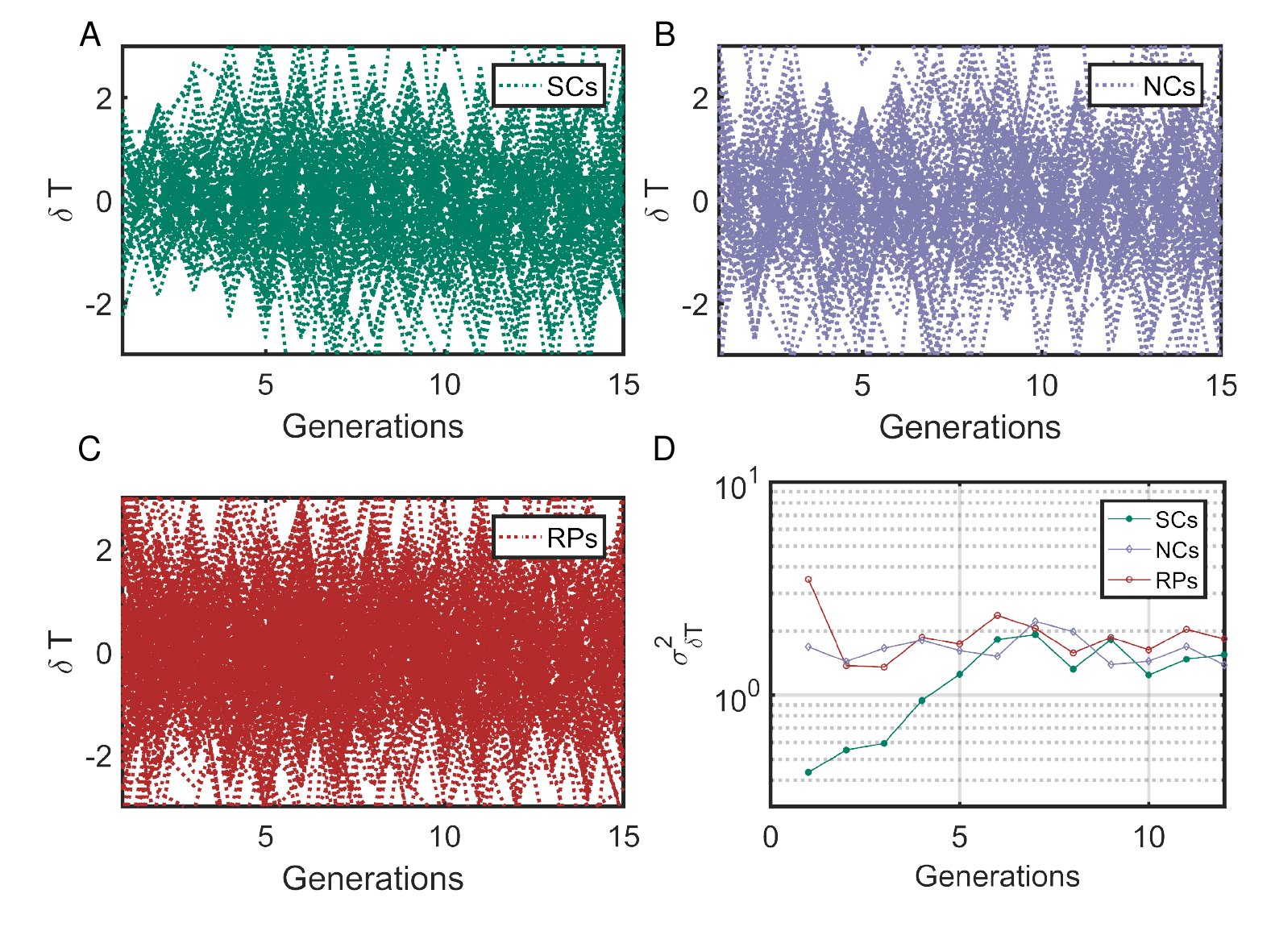}}

\figsupp[Cell size variance ($\sigma_{\delta L_0}^2$) as a function of time.]{Birth size variance $\sigma_{\delta L_0}^2$ was calculated similar to $\sigma_{\delta T}^2$ in Figure3–Figure supplement 1. $\sigma_{\delta L_0}^2$ for SCs increases slowly and saturates at a fixed value after $\sim$7 generations (mean lifetime $\sim$3.5 generations) similar to the time scale observed in the PCF. For NCs with random initial sizes (A), $\sigma_{\delta L_0}^2$ remains constant similar to RPs.  $\sigma_{\delta L_0}^2$ for NCs  with similar birth sizes starts from a value similar to SCs but shoots up to the saturation value within 1 generation.    
}{\includegraphics[width=14cm]{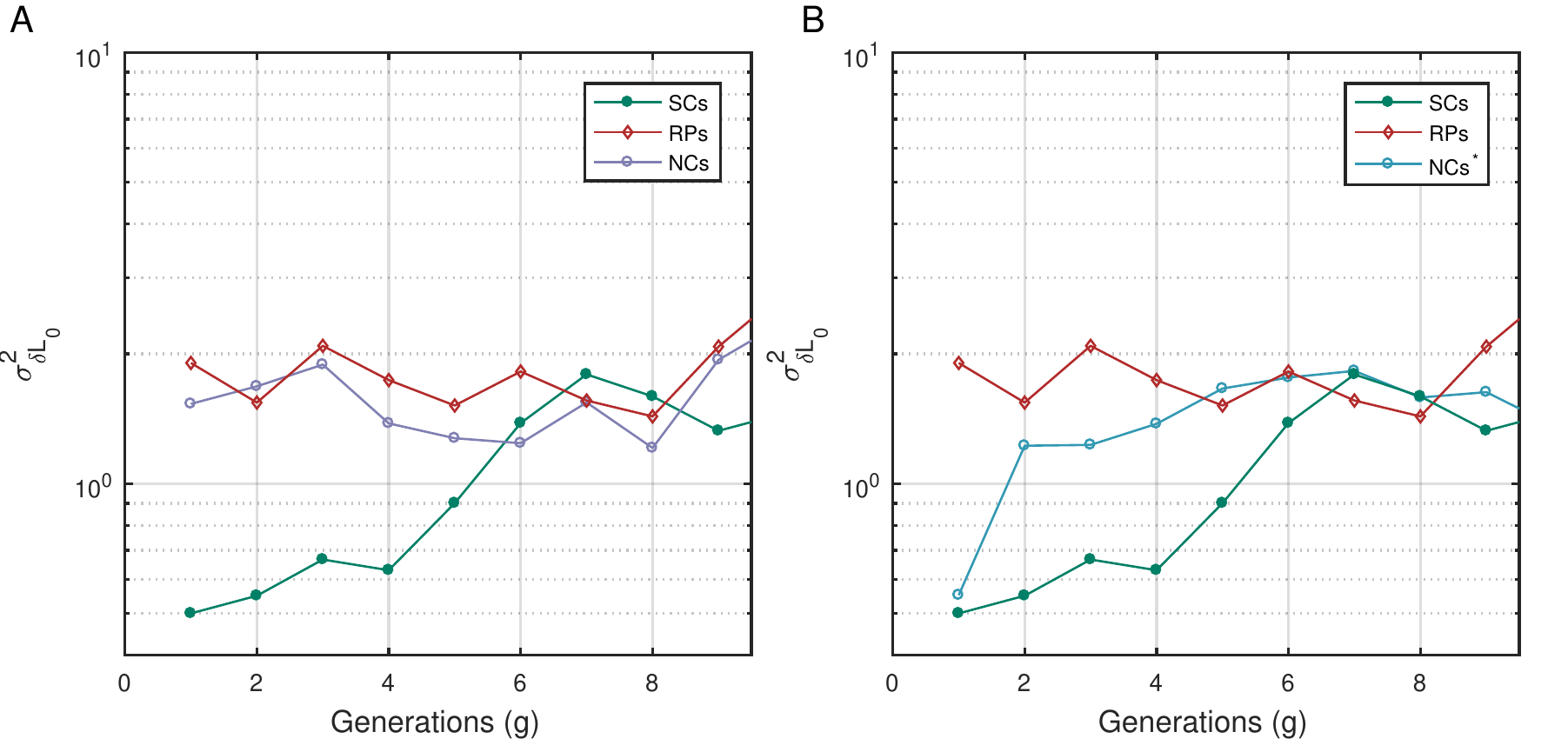}}

\figsupp[Exponential elongation rate difference (${\delta \alpha}$) as a function of time.]{ Individual traces showing the difference between the exponential elongation rates ($\delta \alpha$) for SCs (A), NCs (B), and RPs (C). (D) The mean of $\delta \alpha$ for all cell pairs remain zero along time as expected. For details of $\delta \alpha$ calculations, please refer to the main text. }{\includegraphics[width=14cm]{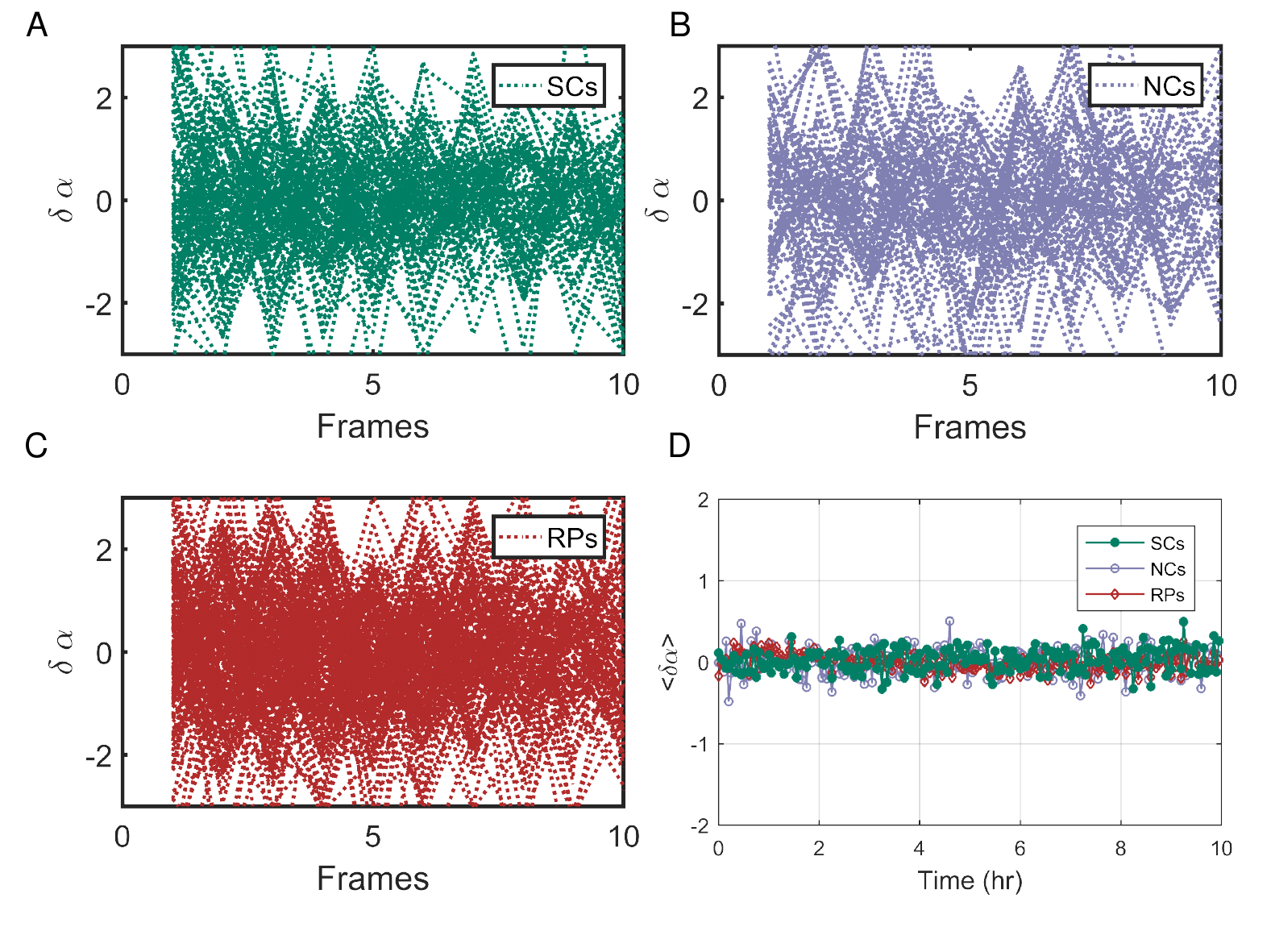}}

\figsupp[Mean fluorescence variance ($\sigma_{\delta f}^2$) as a function of time.]{ Individual traces showing the difference in mean fluorescence intensity ($\delta f$) of gfp expressed in SCs (A), NCs (B), and RPs (C). (D) The variance ($\sigma_{\delta f}^2$ calculated similarly to $\sigma_{\delta T}^2$ in Figure3–Figure supplement 1) of GFP expressed under the control of the Lac Operon promoter in lactose medium (metabolically relevant) is compared with that of GFP expressed under the control of the $\lambda'$ Pr promoter in LB medium (metabolically irrelevant). It is clear that both exhibit no significant difference and a very short memory ($\leq$2 generations).}{\includegraphics[width=14cm]{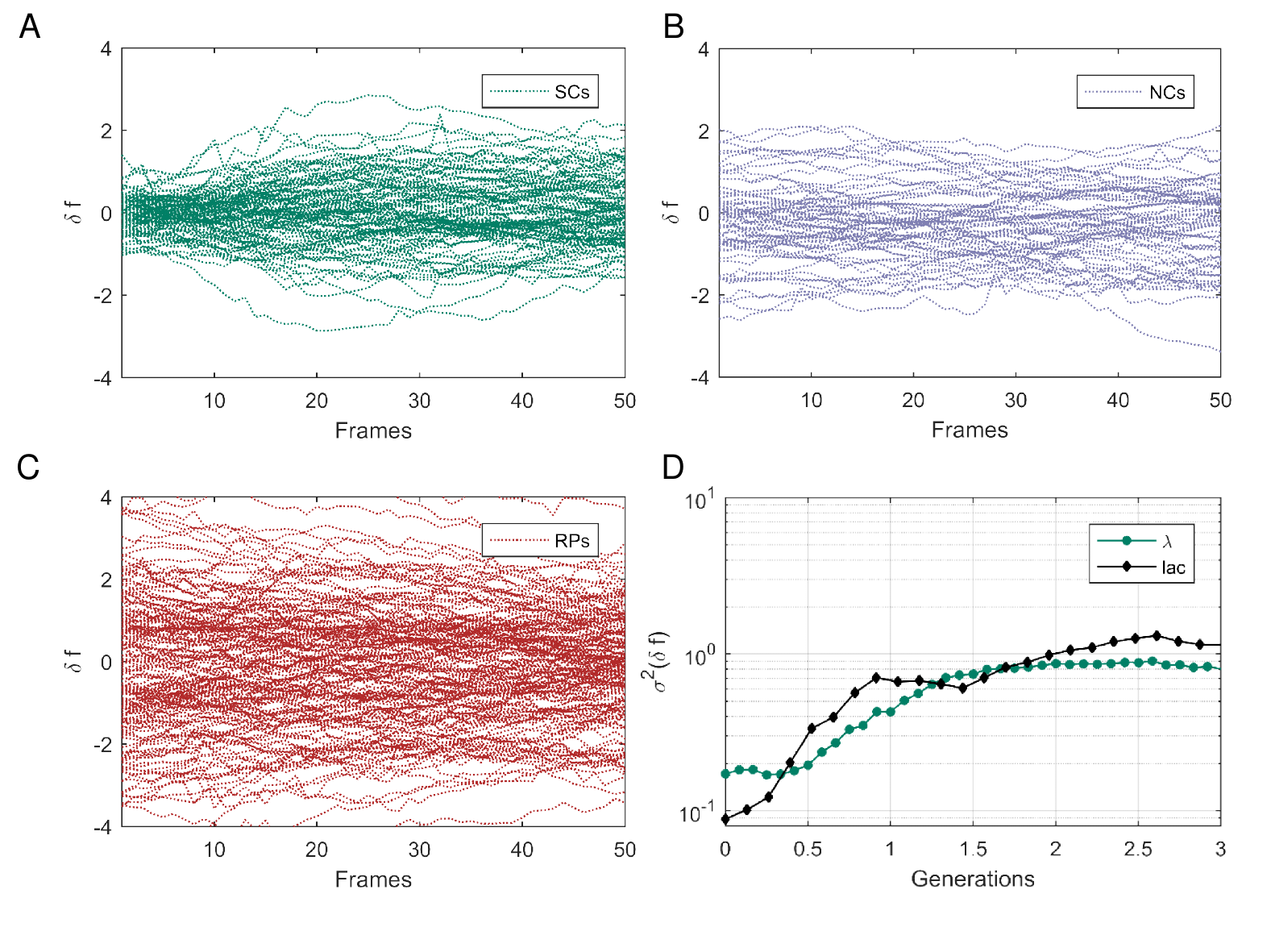}}
\end{figure}

We have also examined how the protein concentration varies over time between the two cells by measuring the concentration of GFP (green fluorescent protein), via its fluorescence intensity, expressed from a constitutive promoter in a medium copy-number plasmid. The variance of fluorescence intensity difference between cell pairs $\delta f$ was calculated as for the growth rate (see Figure3–Figure supplement 4 for details). Upon division, soluble proteins are partitioned symmetrically with both daughters receiving almost the same protein concentration. As expected, $\sigma_{\delta f_{SCs}}^2$ starts from $\sim$0 initially, and diverges to reach saturation within 2 generations (Figure 3C). On the other hand, NCs and RPs exhibit constant variance throughout the whole time, with $\sigma_{\delta f_{RPs}}^2$ twice as large as $\sigma_{\delta f_{NCs}}^2$, which reflects the influence of the shared environment resulting in additional correlations  between NCs. The relatively short-term memory in protein concentration, may be protein specific (Figure3–Figure supplement 4), or it could reflect the fact that in this case the protein is expressed from a plasmid. Nevertheless, this result indicates that cellular properties are controlled differently and can exhibit distinct memory patterns. It is important therefore to distinguish between different cellular characteristics and to examine their inheritance patterns individually.


\section{Discussion}
There has been a rising interest over the past two decades in understanding the contribution of epigenetic factors to cellular properties and their evolution over time. Here, we introduce a new measurement technique that can separate environmental fluctuations from cellular processes. This allows for quantitative measurement of non-genetic memory in bacteria, and reveals its contribution to restraining the variability of cellular properties. Our results show that the restraining force dynamics vary significantly among different cellular properties, and its effects can extend up to $\sim10$ generations. In addition, the growth rate variation emphasizes the effect of division asymmetry, which can help in understanding the mechanism that controls cellular growth rate. The slow increase in the growth rate variance that follows, reflects the effect of inheritance. Since both cells inherit similar content, which ultimately determines the rate of all biochemical activities in the cell and thus its growth rate, it is expected that both cells would exhibit similar growth rates once they make up for the uneven partitioning of size acquired during division. The short memory we see in the protein concentration on the other hand, suggests that cells are less restrictive of their protein concentration. This might be protein specific, or for proteins that are expressed from plasmids only. Nevertheless, these results highlight the importance of such studies, and how this new method can help answer fundamental questions about non-genetic memory and variability in cellular properties.

Finally, in order to understand and characterize the evolution of population growth rate as it reflects its fitness, there is a need to incorporate inheritance effects, which has been thus far assumed to be short lived. This study confirms that cellular memory can persist for several generations, and therefore limits the variation in certain cellular characteristics, including growth rate. Such memory should be considered in future studies and has the potential of changing our perception of population growth and fitness.


\section{Methods and Materials}

\subsection{Key resources table}
\begin{tabular}{| m{2cm} | m{2cm}| m{2cm} | m{2cm} | m{3cm} | }
\toprule
\textbf{Reagent type (species) or resource}  & \textbf{Designation}        & \textbf{Source or reference} &  \textbf{Identifiers}    & \textbf{Additional information} \\
\midrule
strain, strain background (Escherichia coli)     & MG1655          &  Coli Genetic Stock Center (CGSC) & 6300 & F-, $\lambda-$, rph-1  \\
\hline
recombinant DNA reagent	 & pZA3R-GFP	& Lutz and Bujard (1997)	& \href{https://academic.oup.com/nar/article/25/6/1203/1197243 }{https://acad emic.oup. com/ nar/article/ 25/6/1203/ 1197243 }
& GFP expressed from the $\lambda$ Pr promoter
 \\
\hline
recombinant DNA reagent	 & pZA32wt-GFP	&Lutz and Bujard (1997)	& \href{https://academic.oup.com/nar/article/25/6/1203/1197243 }{https://acad emic.oup. com/ nar/article/ 25/6/1203/ 1197243 }
& GFP expressed from the LacO promoter \\
\hline
software, algorithm &	MATLAB  &	MathWorks	& N/A & 	   \\
\hline
software, algorithm	& Oufti	& Paintdakhi et al. (2016)	& \href{http://oufti.org/}{http://oufti .org/} &  \\
\bottomrule
\end{tabular}

\subsection{Device fabrication}
The master mold of the microfluidic device was fabricated in two layers. Initially, the growth channels for the cells were printed on a 1mm x 1 mm fused silica substrate using Nanoscribe Photonic professional (GT). The second layer, containing the main flow channels that supply nutrients and wash out excess cells, was formed using standard soft lithography techniques (\cite{Jenkins2013}; \cite{RodrigoMartinez-DuarteandMarcJ.Madou2016}). SU8 2015 photoresist (MicroChem, Newton, MA) was spin coated onto the substrate to achieve a layer thickness of 30 $\mu$m and cured using maskless aligner MLA100 (Heidelberg Instruments). Following a wash step with SU8 developer, the master mold was baked and salinized. The experimental setup described in the main text was then prepared using this master mold, from PDMS prepolymer and its curing agent (Sylgard 184, Dow Corning) as described in previous studies.

\subsection{Cell culture preparation}
The wild type MG1655 E. coli bacteria were used in all experiments described. Protein content was measured through the fluorescence intensity of green fluorescent protein (GFP) inserted into the bacteria on the medium copy number plasmid pZA (\cite{Lutz1997}). The expression of GFP was controlled by one of two different promoters, the Lac Operon (LacO) promoter was used to measure the expression level of a metabolically relevant protein, while the viral $\lambda$-phage Pr promoter was used to measure the expression level of a constitutive metabolically irrelevant protein.

Two testing media were used in our experiments. M9 minimal medium supplemented with 1g/l casamino acids and 4g/l lactose (M9CL) was used for measuring the expression level from the LacO Promoter, and LB medium was used for all other experiments. The cultures were grown over night at 30$^{\circ}$ C, in either LB or M9CL medium depending on the intended conditions. The following day, the cells were diluted in the same medium and regrown to early exponential phase, Optical Density (OD) between 0.1 and 0.2. When the cells reached the desired OD, they were concentrated into fresh testing medium to an OD$\sim$0.3, and loaded into a microfluidic device. Once enough cells were trapped in the channels, fresh testing medium was pumped through the wide channels of the device to supply the trapped cells with nutrients and wash out extra cells that are pushed out of the channels. The cells were allowed to grow in this device for days, while maintaining the temperature, using a microscope top incubator (Okolab, H201-1-T-UNIT-BL). 

\subsection{Image acquisition, and data analysis}

Images of the channels were acquired every 3 minutes (in LB medium) or 7 minutes (in M9CL medium) in DIC and fluorescence modes using a Nikon eclipse Ti2 microscope with a 100x objective. The size and protein content of the sister cells were measured from these images using the image analysis software Oufti (\cite{Paintdakhi2016}). The data were then used to generate traces such as in Figure 1D, and for further analysis as detailed in the main text. Single-cell measurements were analyzed using MATLAB. Sample autocorrelation functions, Pearson correlation coefficients, sample distributions and curve fitting were all calculated by their implementations in MATLAB toolboxes.

\section{Acknowledgments}
We thank Naama Brenner for helpful discussions and comments on the manuscript. This work was supported in part by the US-Israel Binational Science Foundation.

\bibliography{Sister_reference}
\newpage
\begin{appendixbox}
\section{Mathematical framework} 
Assuming that $x(t)$ is a measurable cellular property, such as cells size, or growth rate, etc. We can present it as:
\begin{align*}x(t)=\Bar{x}+\delta x(t) \end{align*}
Where $\Bar{x}$ is the average of $x(t)$ over time, and $\delta x$ is its fluctuations around $\Bar{x}$. The difference of this measured property between two cells:
\begin{align*} \Delta x(t)= x_1(t) -x_2(t)\end{align*}
Where 1 and 2 represent the two different cells, will average to zero, i.e. $<\Delta x>= 0$. Its variance on the other hand will be: 
\begin{align*} 
\sigma_{\Delta x}^2(t)=<{\Delta x(t)}^2>-<{\Delta x(t)}>^2=2<\delta{x^2(t)}>-2<\delta x_1(t)\delta x_2(t)> 
\end{align*}
Where $<\delta{x^2}>$=$<\delta{x_1}^2>$=$<\delta{x_2}^2>$ is the variance of $x$, which is the same for all cells, and $<\delta x_1(t)\delta x_2(t)>$  is the covariance of the fluctuations in both cells, which when normalized by $\sigma_{\delta x_1}$. $\sigma_{ \delta x_2}$ would give the correlation, i.e. the Pearson Correlation Function (PCF), between the two variables.
On the other hand, if we assume that $x$ is determined by two factors, internal cellular composition ($I(t)$) and external environmental conditions ($E(t)$), such that:
\begin{align*} x(t)=I(t)+E(t)\end{align*}
Then $\sigma_{\Delta x}^2(t)=<[(I_1-I_2)+(E_1-E_2)]^2>$ would depend on whether the two cells share the same environment and/or the same cellular compositions. Therefore, random pair of cells (RPs), which reside in different channels and thus do not share neither the environment nor the internal composition would exhibit a variance:
\begin{align*}RPs: \sigma_{\Delta x}^2(t)=2\sigma_{I}^2+2\sigma_{E}^2+4cov(I,E)\end{align*}
Where $\sigma_{I}^2=<I^2>-<I>^2$ is the variance in the internal composition of the cell (similar for all cells and constant over time), $\sigma_{E}^2=<E^2>-<E>^2$ is the variance in the environmental conditions (also the same for all cells in the same experiment), and $cov(I,E)$, is the covariance of the environment and the internal composition of the cell, which as discussed earlier can influence each other in a trap-specific manner. However, averaging many measurements from different traps erases this effect as clear from Figure 1-Figure supplement 1B (see also \cite{Susman2018}). 
On the other hand, for cells that share the environment but not their internal composition, i.e. neighboring cells (NCs), the variance would be:
\begin{align*}NCs: \sigma_{\Delta x}^2(t)=2\sigma_{I}^2\end{align*}
Note that when the NCs are chosen to have similar size and divide simultaneously at time zero, this variance for cell size would be small initially and its increase would not be constrained by the epigenetic similarity between the two cells as in the case of sister cells (SCs).
And finally, for SCs, which share both the environment and their internal composition, which means that $I_1$ and $I_2$ can be correlated, then:
\begin{align*}SCs: \sigma_{\Delta x}^2(t)=2\sigma_{I}^2-2cov(I_1,I_2)\end{align*}
Where $cov(I_1,I_2)$ is the covariance of the internal states of the cells as a function of time, i.e. the non-genetic memory of the cell. 
Using the definitions above, it is easy to see the relationship between the variance and the PCF. It is also clear that the difference between NCs and RPs variances would provide the contribution of the environment, while the difference between SCs and NCs variances would give the contribution of the internal composition of the cell to the variance, or the epigenetic memory.

\end{appendixbox}

\section{Supplementary material}

\subsection{PCF and error calculation}
The PCF was calculated using following equation:
\begin{equation}
PCF^{(y)}(t)= \frac{1}{\sigma_{y^{(1)}} \sigma_{y^{(2)}}}\sum_{i=1}^{n} (y_i^{(1)} (t)-<y^{(1) } >).(y_i^{(2)} (t)-<y^{(2) } >)
\end{equation}
and the standard deviation(\cite{Bowley1928}):

\begin{equation}
\sigma_{PCF}=\frac{(1-PCF^2)}{\sqrt {n}}
\end{equation}

Where n is the number of cell pairs considered in the calculation.
\newline
\\
\\
\\
\begin{table}[h]
\caption{\label{tab:example}The calculated values of the PCF for SCs were verified by calculating the slopes of best fits to the plots of TimeA vs TimeB graphs (Figure 2- figure supplement 2).}
\begin{tabular}{s l r}
\toprule
Generation &     PCF $\pm \sigma_{PCF}$  & Slope of best fit line ( Figure2–Figure supplement2)     \\
\midrule
1_{st}     & 0.86 $\pm 0.02$          & 0.87  \\
2_{nd}     & 0.65 $\pm 0.05$         & 0.69  \\
3_{rd}     & 0.54 $\pm 0.06$         & 0.44  \\
4_{th}     & 0.36 $\pm 0.07$         & 0.42  \\
5_{th}     & 0.28 $\pm 0.08$         & 0.25  \\
6_{th}     & 0.23 $\pm 0.08$         & 0.25  \\
7_{th}     & 0.12 $\pm 0.09$        & 0.11  \\
8_{th}     & 0.23 $\pm 0.09$          & 0.25  \\
9_{th}     & 0.00 $\pm 0.09$         & 0.00  \\

\bottomrule
\end{tabular}

\end{table}

\end{document}